\documentclass[twocolumn,showpacs,floatfix,prl]{revtex4}

\usepackage[dvips]{epsfig}

\usepackage{float}

\begin{document}

\title{Topological Insulators in Three Dimensions}

\author{Liang Fu, C.L. Kane and E.J. Mele}
\affiliation{Dept. of Physics and Astronomy, University of Pennsylvania,
Philadelphia, PA 19104}

\begin{abstract}
We study three dimensional generalizations of the quantum spin
Hall (QSH) effect.  Unlike two dimensions, where the QSH
effect is distinguished by a single $Z_2$ topological invariant,
in three dimensions there are 4
invariants distinguishing 16 ``topological insulator" phases.
There are two general classes:
weak (WTI) and strong (STI) topological insulators.
The WTI states are equivalent to layered 2D QSH states, but are
fragile because disorder continuously connects them to
band insulators.  The STI states are robust
and have surface states that realize the
2+1 dimensional parity anomaly without fermion doubling,
giving rise to a novel ``topological metal" surface phase.
We introduce a tight binding model
which realizes both the WTI and STI phases, and we discuss the
relevance of this model to real three dimensional materials, including
bismuth.

\end{abstract}

\maketitle

In recent years, the advent of spintronics has motivated the study of
the effects of spin orbit interactions (SOI) on the electronic
structure of solids.  SOI leads to the spin Hall
effect\cite{murakami1,sinova}, which has been observed in
GaAs\cite{kato,wunderlich}.  We proposed\cite{km1}
 that in graphene the SOI leads to the quantum spin
Hall (QSH) effect.  In the QSH phase there is a bulk excitation gap
along with gapless spin-filtered edge states.  The QSH phase is
distinguished from a band insulator by a $Z_2$ topological
invariant\cite{km2}, which is a generalization applicable to time
reversal invariant systems of the TKKN invariant of the integer
quantum Hall effect\cite{tknn}.  Because of the weak SOI in carbon,
the SOI induced energy gap in graphene is likely to be quite
small\cite{weakso}.  However, Murakami has recently suggested that
bismuth bilayers may provide an alternative venue for the QSH
effect\cite{murakami2}.  This breakthrough provides a new direction
for the experimental observation of this phase.

In this paper we consider the generalization of the QSH effect to
three dimensions (3D).  Our work builds on recent progress by Moore and
Balents\cite{moore}, who showed that time reversal invariant energy
bands in 3D are characterized by four $Z_2$ invariants,
leading to 16 classes of ``topological insulators". A similar result
has also been obtained by Roy\cite{roy}. Here, we will explain the
physical meaning of these invariants and characterize the phases they
distinguish. One of the four invariants is of special significance
and distinguishes what we will refer to as ``weak" and ``strong"
topological insulators. With disorder, the weak topological insulator
(WTI) is equivalent to a band insulator, while the strong topological
insulator (STI) remains robust.  We show WTIs and STIs have surface
states with an even and odd number of Dirac points respectively.  The
latter case leads to a new ``topological metal" surface phase, which
we characterize.   We introduce a tight binding model on a distorted
diamond lattice, which realizes both the WTI and STI phases, allowing
the surface states to be studied explicitly.
%While the diamond
%lattice model is not designed to describe any specific material, we
%argue that it provides an organizing principle for the rational
%search for topological insulating materials.

In Ref. \onlinecite{fu} we established the connection between the
$Z_2$ invariant for the bulk QSH phase and the spin filtered edge
states. We begin by reviewing that argument in a way which makes the
generalization to 3D transparent. The 2D invariant can
be understood using a Laughlin type construction\cite{laughlin} on a
cylinder threaded by magnetic flux $\Phi = 0$ or $\pi$ (in units of
$\hbar/e$).  The invariant characterizes the change in the {\it time
reversal polarization} (TRP), which signals the presence of a Kramers
degeneracy at the ends, when $\Phi$ is changed from $0$ to $\pi$.  If
the cylinder consists of a single unit cell in the circumferential
($x$) direction, then the magnetic flux threading the cylinder plays
the role of the crystal momentum $k_x$ in band theory.
The spectrum of the discrete end states of the cylinder as a function
of flux then reflects the edge state spectrum as a function of
momentum.   The change in
the TRP as a function of flux determines the way the Kramers
degenerate end states at the edge time reversal invariant momenta
(TRIM) $k_x =\Lambda_1 = 0$ and $k_x=\Lambda_2 = \pi$
are connected to each other.  In the QSH phase
the Kramers pairs ``switch partners" (Fig. 1a), reflecting the
change in the TRP, while in the conventional insulator (Fig. 1b) they do not.
It follows that in the
QSH phase edge states traverse the bulk energy gap, and cross
the Fermi energy an {\it odd} number of times between $\Lambda_a$ and
$\Lambda_b$.
In the insulating phase, the edge states cross the
Fermi energy an even number of times if at all.  They are not topologically
protected, since changes in the Hamiltonian at the edge
can push the entire edge band out of the bulk gap.

\begin{figure}
 \centerline{ \epsfig{figure=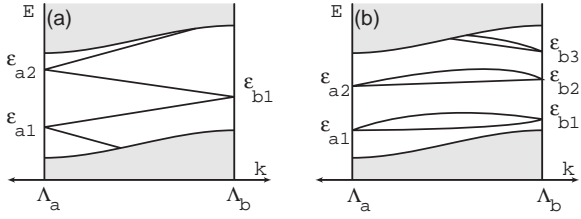,width=3in} }
 \caption{Schematic surface (or edge) state spectra as a function of
 momentum along a line connecting $\Lambda_{a}$ to $\Lambda_{b}$ for
 (a) $\pi_a\pi_b=-1$ and (b) $\pi_a\pi_b=+1$.  The shaded region shows the
  bulk states.  In (a) the TRP changes
 between $\Lambda_a$ and $\Lambda_b$, while in (b) it does not.
 }
 \label{surface}
\end{figure}

The change in the TRP between $\Lambda_1$
and $\Lambda_2$ is related to the
bulk band structure, defined for a 2D system with periodic
boundary conditions in both directions.  The 2D
Brillouin zone has four TRIM, $\Gamma_i$, which are related to $-\Gamma_i$
by a reciprocal lattice vector.   For an edge perpendicular to ${\bf G}$
the 1D edge TRIM $\Lambda_{a=1,2}$ are projections of pairs
$\Gamma_{i=a1}$, $\Gamma_{i=a2}$, which satisfy
$\Gamma_{a1}-\Gamma_{a2}= {\bf G}/2$, onto the line perpendicular to
${\bf G}$.

The TRP associated with $\Lambda_a$
can be expressed as
$\pi_a = \delta_{i=a1} \delta_{i=a2}$, where\cite{fu}
\begin{equation}
\delta_i = \sqrt{{\rm Det}[w(\Gamma_i)]}/{\rm Pf}[w(\Gamma_i)] =
\pm 1.  \label{delta}
\end{equation}
Here the unitary matrix $w_{ij}({\bf k}) = \langle u_i(-{\bf
k})|\Theta|u_j({\bf k)}\rangle$.   At ${\bf k} = \Gamma_i$,
$w_{ij}=-w_{ji}$, so the Pfaffian ${\rm Pf}[w]$ is
defined.  $\pi_a$ is free of the ambiguity of the square root in
(\ref{delta}), provided the square root is chosen continuously
as a function of ${\bf k}$.
 However,
$\pi_{a}$ is not gauge invariant.  A ${\bf k}$ dependent gauge
transformation can change the sign of any pair of $\delta_i$'s.
This reflects the physical fact that the end Kramers
degeneracy depends on how the crystal is terminated.  It is
similar to the ambiguity of the charge polarization\cite{fu}.
The
product, $\pi_1\pi_2 = \delta_1\delta_2\delta_3\delta_4$, {\it is} gauge
invariant, and characterizes the {\it change} in TRP
 due to changing the flux from
$\Lambda_1 = 0$ to $\Lambda_2=\pi$.  This
defines the single $Z_2$ invariant in 2D, and using
the above argument, determines the connectivity of the edge state
spectrum.

In 3D there are 8 distinct TRIM, which are expressed in
terms of primitive reciprocal lattice vectors as  $\Gamma_{i=(n_1 n_2
n_3)} = (n_1 {\bf b}_1 + n_2 {\bf b}_2 + n_3 {\bf b}_3)/2$, with $n_j
= 0,1$.  They can be visualized as the vertices of a cube as in Fig.
2.  A gauge transformation can change the signs of $\delta_i$
associated with
any four $\Gamma_i$
that lie in the same plane.  Modulo these gauge transformations,
there are 16 invariant configurations of $\delta_i$. These can be
distinguished by 4 $Z_2$ indices $\nu_0; (\nu_1\nu_2\nu_3)$, which we
define as
\begin{eqnarray}
(-1)^{\nu_0} &=& \prod_{n_j = 0,1} \delta_{n_1n_2n_3} \\
(-1)^{\nu_{i=1,2,3}} &=& \prod_{n_{j\ne i} = 0,1; n_i = 1} \delta_{n_1 n_2
n_3}.
\end{eqnarray}
$\nu_0$ is independent of the choice of ${\bf b}_k$.  $(\nu_1\nu_2\nu_3)$
are not, but they can be identified with
${\bf G}_\nu \equiv \sum_i \nu_i {\bf b}_i$, which belongs to
the 8 element {\it mod 2 reciprocal lattice}, in which vectors that
differ by $2{\bf G}$ are identified.  $(\nu_1\nu_2\nu_3)$ can be
interpreted as Miller indices for ${\bf G}_\nu$.

$\nu_{0-4}$ are equivalent to the four invariants introduced by
Moore and Balents using general homotopy arguments.  The power of
the present approach is that it allows us to characterize the
surface states on an arbitrary crystal face.  Generalizing the
Laughlin argument to 3D, consider a system with open
ends in one direction and periodic boundary conditions in the
other two directions.  This can be visualized as a torus with a
finite thickness (a ``Corbino donut"), which has an inside and an
outside surface.  Viewed as a 1D system, we then seek
to classify the changes in the Kramers degeneracy associated with
the surfaces as a function of two fluxes threading the torus (or
equivalently as a function of the two components of the surface
crystal momentum).

For a surface perpendicular to ${\bf G}$, the
surface Brillouin zone has four TRIM $\Lambda_a$ which are
the projections of pairs  $\Gamma_{a1}$, $\Gamma_{a2}$,
that differ by ${\bf G}/2$, into the plane perpendicular to ${\bf G}$.
Due to Kramers' degeneracy, the surface
spectrum has two dimensional {\it Dirac points} at $\Lambda_a$.
The {\it relative} values of $\pi_a = \delta_{a1}\delta_{a2}$
determine how these Dirac points are connected to one
another, as illustrated in Fig. 1.  For any path connecting
$\Lambda_a$ to $\Lambda_b$, the surface band structure
will resemble Fig. 1a (1b) for $\pi_a \pi_b = -1 (+1)$, and
the surface bands will intersect the Fermi energy an odd (even)
number of times.  It follows that the surface Fermi arc
divides the surface Brillouin zone into two regions.  The
Dirac points at the TRIM
$\Lambda_a$ with $\pi_a=+1$ are on one side, while those with $\pi_a=-1$
are on the other side.

\begin{figure}
 \centerline{ \epsfig{figure=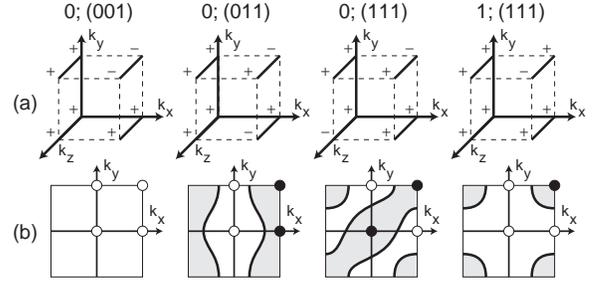,width=3in} }
 \caption{Diagrams depicting four different phases indexed by $\nu_0
 ; (\nu_1\nu_2\nu_3)$.  (a) depicts $\delta_i$ at
 the TRIM $\Gamma_i$ at the vertices of the cube.
 (b) characterizes the
 $001$ surface in each phase.  The surface TRIM $\Lambda_a$ are
 denoted by open (filled) circles for $\pi_a = \delta_{a1}\delta_{a2}
 = +1 (-1)$.  They are projections of $\Gamma_{a1}$ and $\Gamma_{a2}$, which are
 connected by solid lines in (a).
The thick lines and shaded regions in (b) indicate possible
surface Fermi arcs which enclose specific $\Lambda_a$.}
 \label{phase}
\end{figure}

In Fig. \ref{phase} we depict $\delta_i$ for four different
topological classes, along with the predictions for the edge state
spectrum for a 001 face. The surface Fermi arc encloses
either $0 (4)$, $1 (3)$, or $2$ Dirac points. When the
number of Dirac points is not $0 (4)$, there {\it must} be surface
states which connect the bulk conduction and valence bands.
% As in 2D\cite{km1}, these surface states are ``spin
%filtered" in that the expectation value of the spin is correlated
%with the propagation direction, such that $\langle {\bf s}(-{\bf
%k})\rangle = - \langle {\bf s}({\bf k})\rangle$.  Spin density and
%charge current are thus coupled.

There are two classes of phases depending on the parity of $\nu_0$.
For $\nu_0 = 0$ each face has either $0 (4)$ or $2$ enclosed Dirac points. For
a face ${\bf G}=\sum_i m_i {\bf b}_i$ there are $0 (4)$ Dirac points
for $m_i = \nu_i\ {\rm mod}\ 2$ ($i=1,2,3$) and 2 Dirac points
otherwise.  These phases can be interpreted as layers of 2D QSH
states stacked in the ${\bf G}_\nu$
direction.  They resemble three dimensional quantum Hall
phases\cite{kohmoto}, which are indexed by a triad of Chern integers
that define a reciprocal lattice vector ${\bf G}$ perpendicular to
the layers and give the conductivity
$\sigma_{ij}=(e^2/h)\varepsilon_{ijk} G_k/(2\pi)$. In the present case, ${\bf
G}_\nu$ is defined modulo $2{\bf G}$, so that layered QSH phases stacked
along ${\bf G}_\nu$ and ${\bf G}_\nu + 2 {\bf G}$ are equivalent.

The presence or absence of surface states in the $\nu_0=0$ phases is
delicate.  For the 0;(001) phase in Fig. 2, the 100 face has two Dirac
points, while the 801 face has 0(4).  This sensitivity
%to the precise surface orientation
is a symptom of the fact that the topological
distinction of these phases relies on the translational symmetry
of the lattice.  Indeed, if the unit cell is doubled, the two Dirac
points fold back on one another.  A weak periodic
potential then opens a gap.  It is thus likely that disorder will
eliminate the topological distinction between these phases and simple
insulators.  Surface states will generically be localized.
For this reason, we refer to the $\nu_0=0$ phases as
``weak" topological insulators.  Nonetheless,
the weak invariants have important implications for clean
surfaces.

The $\nu_0=1$ phases are more robust, and we refer
to them as ``strong" topological insulators.
In this case the surface Fermi arc encloses  1(3) Dirac points
on {\it all} faces.  If the Fermi energy
is exactly at the Dirac point
this provides a time reversal invariant realization
of the 2+1 dimensional parity anomaly\cite{jackiw,semenoff,fradkin,haldane}
{\it without fermion doubling}.  This can occur because the Dirac
point partners reside on opposite surfaces.
 For a generic
Fermi energy inside the bulk gap the surface Fermi arc will
enclose a single Dirac point.  This defines a two
dimensional ``topological metal" that is topologically protected
because a quantized Berry's phase of $\pi$ is acquired by an electron circling
the Fermi arc.  This
Berry's phase implies that with disorder the surface is in the
symplectic universality class\cite{hikami,suzuura}, which is not localized by weak
disorder.

\begin{figure}
 \centerline{ \epsfig{figure=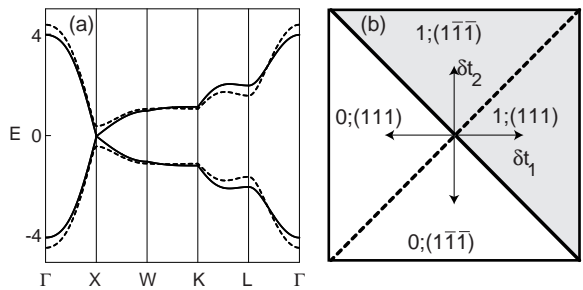,width=3in} }
 \caption{Energy bands for (a) the model
 (\ref{tbmodel}) with $t=1$, $\lambda_{SO}= .125$.  The symmetry
 points are $\Gamma = (0,0,0)$, $X = (1,0,0)$, $W =
 (1,1/2,0)$, $K=(3/4,3/4,0)$ and $L = (1/2,1/2,1/2)$ in units of
 $2\pi/a$.  The dashed
 line shows the energy gap due to $\delta t_1 = .4$.  (b)
 shows the phase diagram as a function of $\delta t_1$ and
 $\delta t_2$ (for bonds in the $111$ and
 $1\bar 1\bar 1$ directions) with phases indexed according to
 cubic Miller indices for ${\bf G}_\nu$.
 The shaded region is the STI phase.}
\end{figure}

To develop an explicit model of these phases we
consider a 4 band tight binding model of $s$ states on a diamond
lattice with SOI, that
generalizes the 2D honeycomb lattice model\cite{semenoff,haldane,km1}.
\begin{equation}
H=t\sum_{\langle ij\rangle} c_i^\dagger c_j
+ i(8\lambda_{SO}/a^2) \sum_{\langle\langle ij \rangle\rangle}
 c_i^\dagger {\bf s} \cdot ({\bf d}_{ij}^1 \times {\bf d}_{ij}^2) c_j. \\
 \label{tbmodel}
\end{equation}
The first term is a nearest neighbor hopping term connecting the
two fcc sublattices of the diamond lattice.  The second term connects
second neighbors with a spin dependent amplitude.  ${\bf d}_{ij}^{1,2}$ are
the two nearest neighbor bond vectors traversed between sites $i$ and $j$.
$a$ is the cubic cell size.  The energy bands
are shown in Fig. 3a.  Due to inversion symmetry, each band is doubly degenerate.
The conduction and valence
bands meet at 3D Dirac points at the three inequivalent X points
$X^r = 2\pi\hat r/a$, where $r = x, y, z$.
This degeneracy is lifted by symmetry lowering modulations of the four
nearest neighbor bonds $t \rightarrow t+\delta t_p$, with
$p=1,...,4$.

Near $X^z$ the low
energy effective mass model has the form of a 3+1 dimensional Dirac
equation,
\begin{equation}
{\cal H}_{\rm eff}^z = ta \sigma^y q_z + {4\lambda_{SO}a} \sigma^z (s^x
q_x - s^y q_y) + m^z \sigma^x.
\end{equation}
Here ${\bf q} = {\bf k} - X^z$ and
$m^z = \sum_p \delta t_p {\rm sgn}[{\bf d}_p\cdot \hat z]$.
${\bf d}_p$ is the bond vector associated with the $p$'th nearest
neighbor bond.
The Pauli matrices $\sigma^i$ are associated with the sublattice
degree of freedom, while $s^i$ describe the spin.
${\cal H}_{\rm eff}^{x,y}$ are the same
with $x$, $y$ and $z$
permuted in $q_i$ and $s^i$, but not $\sigma^i$.  Transitions between
different phases occur when the masses at any of the $X^r$
vanish.  $\delta t_p = 0$ is thus a
multicritical point separating 8 different time reversal invariant phases.

Determining $\nu_i$ for these phases using (1-3) requires
eigenvectors that are defined continuously
throughout the Brillouin zone\cite{fu}.  Since the Chern integers vanish, this
is always possible.  Determining the appropriate phases numerically,
however, is nontrivial.  An alternative (though tedious) numerical
approach would be to characterize the Pfaffian function introduced in
Ref. \onlinecite{km2}.  Generalizing the results of Ref. \onlinecite{fu}
it can be shown that the
product of 4 $\delta_i$ in any plane is related to the zeros of the
Pfaffian in that plane, which can be identified without choosing
phases.  For the present problem, however, we are fortunate because
the eigenvectors can be determined analytically, allowing for the
continuation of $w_{ij}({\bf k})$ between the different $\Gamma_i$.
We find
\begin{equation}
\delta_i = {\rm sgn}[\sum_p (t + \delta t_p) \cos \Gamma_i\cdot
({\bf d}_p-{\bf d}_1)].
\end{equation}
For small $\delta t_p$,
$\delta = 1$ at ${\bf k}=0$ and at 3 of the L points.  $\delta =
-1$ at the 4th L point.  At $X^r$, $\delta = {\rm sgn}
[m^r]$.  When one of the four bonds is weaker than the others ($\delta t_p <0$,
$\delta t_{(p'\ne p)}=0$ for instance) the system is in a WTI
phase, which may be interpreted as a QSH state
layered in the ${\bf d}_p$ direction.  There are 4 such states,
depending on $p$, of which two are shown in Fig. 3b.  We labeled
the phases with the conventional cubic
Miller indices for ${\bf G}_\nu$.
$111$ and $1\bar 1 \bar 1$ are distinct elements of the fcc mod 2 reciprocal
lattice.  When one of the
bonds is stronger than the others the system is in one of four STI
phases.  The band insulator $0;(000)$ is not
perturbatively accessible from
this critical point.  However, in the tight binding model it
occurs when one bond is turned up so that $\delta t_1 > 2 t$.
A staggered sublattice potential also leads to a band
insulator, but the strength must exceed a finite value (set by
$\lambda_{SO}$) before that transition occurs.

\begin{figure}
 \centerline{ \epsfig{figure=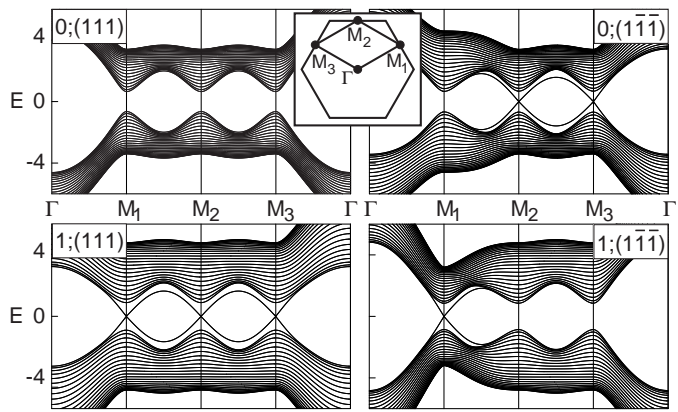,width=3.5in} }
\caption{2D band structures for a slab with a 111 face
for the four phases in Fig. 3.  The states
crossing the bulk energy gap are localized at the surface.  In the
WTI (STI) phases there are an even (odd) number of Dirac points in
the surface spectrum.  The inset shows the surface Brillouin zone. }
\end{figure}

To study the surface states, we solve (\ref{tbmodel}) in a slab geometry.
Fig. 4 shows the 2 dimensional band structures of the four phases in Fig. 3 for a slab
parallel to the $111$ surface along lines that visit each of
the four TRIM.  The plots with the same $\nu_0$ can be
viewed as different faces of the same state. The bulk states above the
bandgap are clearly seen.  In addition, there are surface states
which traverse the gap.  In the WTI phases $0;(111)$ and $0;(1\bar 1
\bar 1)$
there are $0$ and $2$ 2D Dirac points,
on both the top and bottom surfaces, as expected from the
general arguments given above.  In the STI phases $1;(111)$ and $1;(\bar
1\bar 1 1)$ there is $1(3)$ Dirac point on each surface.  In each case,
the non degenerate surface states near the Dirac points are spin filtered, such that
$\langle \vec s(-{\bf k}) \rangle = - \langle \vec s({\bf
k})\rangle$.  Spin density and charge current are thus coupled.

Though the 4 band diamond lattice model is simple, it is
probably not directly relevant to any specific material.  However, it
may give insight into the behavior of real crystals.
Consider a sequence of crystal
structures obtained from diamond by continuously
displacing the fcc sublattices in the (111) direction:
$$
{\rm diamond} \rightarrow {\rm graphite (ABC)} \rightarrow {\rm
cubic}.
$$
Starting with diamond, the 111 nearest neighbor bond is stretched,
leading to the $0;(111)$ WTI phase.  As the
sublattice is displaced further both sublattices eventually reside
in the same plane with a structure similar to ABC stacked
graphite.  Displacing further, the lattice eventually becomes cubic.
At this point, the gap closes, and the system is metallic.
The $s$ state model remains in the WTI phase up to the cubic point.

Bismuth has the rhombohedral A7 structure, which can be viewed as a
cubic lattice distorted ``toward diamond", along with a trigonal
distortion of the fcc Bravais lattice.  Murakami showed that a
bilayer of bismuth, whose structure is similar to a single plane of
graphene, is in the QSH phase.  This suggests that for weak coupling
between bilayers bismuth is in the $0;(111)$ WTI phase.
While this agrees with the simple model presented above, a realistic
description of bismuth requires a theory which incorporates
bismuth's five valence bands\cite{liu}.

It will be interesting to search for materials in the STI phase,
which occur on the ``other side of diamond" in our sequence.  We
hope that the exotic surface properties predicted for this phase will
stimulate further experimental and theoretical efforts.

It is a pleasure to thank Joel Moore and Leon Balents for helpful
discussions.  This work was supported by NSF grants DMR-0079909 and
DMR-0605066 and DOE grant DE-FG02-ER-0145118.

{\it Note added:}  In subsequent work we have predicted that a number of
specific materials are STI's\cite{fk}.  These include
the semiconducting alloy Bi$_{1-x}$Sb$_x$ as well as $\alpha$-Sn and
HgTe under uniaxial strain.

\end{document}